\documentclass[11pt]{article}

\usepackage{times}
\usepackage{float}

\usepackage{amsfonts,amstext,amsmath,amssymb,amsthm}

\usepackage{chngpage}
\usepackage{rotating}
\usepackage{graphicx}
\usepackage{epstopdf} 

\usepackage{caption}
\usepackage{subcaption}
\usepackage{mathrsfs}

\long\def\symbolfootnote[#1]#2{\begingroup%
\def\thefootnote{\fnsymbol{footnote}}\footnote[#1]{#2}\endgroup}

\usepackage{hyperxmp}
\usepackage[%
    bookmarks=true,         
    unicode=false,          
    pdftoolbar=true,        
    pdfmenubar=true,        
    hyperfootnotes=false,   
    pdffitwindow=false,     
    pdfstartview={FitH},    
    pdftitle={Asymptotic Exponentiality of the First Exit Time of the Shiryaev--Roberts Diffusion with Constant Positive Drift},    
    pdfauthor={Aleksey S. Polunchenko},     
    pdfsubject={Asymptotic Exponentiality of the First Exit Time of the Shiryaev--Roberts Diffusion with Constant Positive Drift},   
    pdfcreator={Aleksey S. Polunchenko},   
    pdfkeywords={First exit times,Generalized Shiryaev--Roberts procedure,Laplace transform,Markov diffusions,Moment generating function,Quickest change-point detection,Whittaker functions}, 
    pdfnewwindow=true,      
    colorlinks=false,       
    linkcolor=red,          
    citecolor=green,        
    filecolor=magenta,      
    urlcolor=cyan           
]{hyperref}

\usepackage{suffix}
\usepackage{mathtools}

\usepackage[T1]{fontenc}

\usepackage{titlesec} 

\titleformat{\section}{\large\bfseries\uppercase}{\thesection.}{.5em}{}
\titlespacing*{\section}{0pt}{*3}{*2}
\titleformat{\subsection}{\normalfont\bfseries}{\thesubsection.}{.5em}{}
\titlespacing*{\subsection}{0pt}{*3}{*2}
\titleformat{\subsubsection}{\normalfont\bfseries}{\thesubsubsection.}{.5em}{}
\titlespacing*{\subsubsection}{0pt}{*3}{*2}

\numberwithin{equation}{section}

\usepackage[sort&compress]{natbib}

\DeclarePairedDelimiterX\MeijerM[3]{\lparen}{\rparen}%
{\begin{smallmatrix}#1 \\ #2\end{smallmatrix}\delimsize\vert\,#3}

\newcommand\MeijerG[8][]{%
  G^{\,#2,#3}_{#4,#5}\MeijerM[#1]{#6}{#7}{#8}}

\WithSuffix\newcommand\MeijerG*[7]{%
  G^{\,#1,#2}_{#3,#4}\MeijerM*{#5}{#6}{#7}}

\long\def\symbolfootnote[#1]#2{\begingroup%
\def\thefootnote{\fnsymbol{footnote}}\footnote[#1]{#2}\endgroup}

\renewcommand{\Pr}{\mathbb{P}} 
\DeclareMathOperator{\EV}{\mathbb{E}} 

\DeclareMathOperator{\E1}{E_1}

\renewcommand{\le}{\leqslant} 
\renewcommand{\ge}{\geqslant}

\newcommand{\abs}[1]{\left\vert#1\right\vert}

\theoremstyle{plain} 
\newtheorem{theorem}{Theorem}[section]

\graphicspath{{./gfx/}}

\usepackage[norefs,nocites]{refcheck}
\usepackage{silence}
\begin{document}

\title{\textbf{\Large Asymptotic Exponentiality of the First Exit Time of the Shiryaev--Roberts Diffusion with Constant Positive Drift}}

\date{}
\author{}
\maketitle

\begin{center}
\null\vskip-2cm\author{
\textbf{\large Aleksey\ S.\ Polunchenko}\\
Department of Mathematical Sciences, State University of New York at Binghamton,\\Binghamton, New York, USA
}
\end{center}
%
%
%
%
\symbolfootnote[0]{\normalsize\hspace{-0.6cm}Address correspondence to A.\ S.\ Polunchenko, Department of Mathematical Sciences, State University of New York (SUNY) at Binghamton, 4400 Vestal Parkway East, Binghamton, NY 13902--6000, USA; Tel: +1 (607) 777--6906; Fax: +1 (607) 777--2450; E-mail:~\href{mailto:aleksey@binghamton.edu}{aleksey@binghamton.edu}.}\\
%
%
{\small\noindent\textbf{Abstract:} We consider the first exit time of a Shiryaev--Roberts diffusion with constant positive drift from the interval $[0,A]$ where $A>0$. We show that the moment generating function (Laplace transform) of a suitably standardized version of the first exit time converges to that of the unit-mean exponential distribution as $A\to+\infty$. The proof is explicit in that the moment generating function of the first exit time is first expressed analytically and in a closed form, and then the desired limit as $A\to+\infty$ is evaluated directly. The result is of importance in the area of quickest change-point detection, and its discrete-time counterpart has been previously established---although in a different manner---by Pollak \& Tartakovsky~\citeyearpar{Pollak+Tartakovsky:TPA2009}.
}
\\ \\
%
%
{\small\noindent\textbf{Keywords:} {First exit times; Generalized Shiryaev--Roberts procedure; Laplace transform; Markov diffusion; Moment generating function; Quickest change-point detection; Whittaker functions.}
\\ \\
%
%
%
%
%
%
%
%
%

\section{Introduction} 
\label{sec:intro}

This work centers around the so-called Generalized Shiryaev--Roberts (GSR) stochastic process, a time-homogeneous Markov diffusion well-known in the area of quickest change-point detection. See, e.g.,~\cite{Shiryaev:SMD61,Shiryaev:TPA63,Shiryaev:Book78,Shiryaev:Bachelier2002,Shiryaev:Book2011},~\cite{Pollak+Siegmund:B85},~\cite{Feinberg+Shiryaev:SD2006},~\cite{Burnaev+etal:TPA2009},~\cite{Polunchenko+Sokolov:MCAP2016}, and~\cite{Polunchenko:SA2016,Polunchenko:SA2017}. More specifically, for a reason to be made clear shortly, the case of interest is that of the GSR process with a constant positive drift. Formally, we shall deal with the solution $(R_{t}^r)_{t\ge0}$ of the stochastic differential equation (SDE)
\begin{align}\label{eq:Rt_r-def}
dR_t^r
&=
dt+\mu R_t^r dB_t
\;\text{with}\;
R_0^r\triangleq r\ge0,
\end{align}
where $\mu\neq0$ is a given coefficient (whose meaning is explained below), and $(B_t)_{t\ge0}$ is standard Brownian motion (i.e., $\EV[dB_t]=0$, $\EV[(dB_t)^2]=dt$, and $B_0=0$); the initial value $r$ is often referred to as the headstart. The process $(R_t)_{t\ge0}$ governed by~\eqref{eq:Rt_r-def} is a GSR process with unit drift and headstart $r\ge0$; the unit drift can be trivially adjusted to any other constant positive level. The main contribution of this work concerns the distribution of the first exit time of $(R_{t}^r)_{t\ge0}$ from the interval $[0,A]$ with $A>0$ given, i.e., the stopping time:
\begin{align}\label{eq:T-GSR-def}
\mathcal{S}_A^r
&\triangleq
\inf\{t\ge0\colon R_{t}^r=A\}\;\text{such that}\;\inf\{\varnothing\}=+\infty,
\end{align}
where $A>0$ is a preset level. Correspondingly, the process $(R_{t}^r)_{t\ge0}$ and its characteristics pose interest only up to the point of ``extinction'' at time instance $\mathcal{S}_A^r$, i.e., conditional on $\{\mathcal{S}_A^r>t\}$ for a given $t\ge0$.

Just as does the GSR diffusion $(R_{t}^r)_{t\ge0}$---whether with constant positive drift or with a more general affine drift---the stopping time $\mathcal{S}_A^r$, too, plays a major role in the theory of quickest change-point detection: it is the Run Length of the so-called Generalized Shiryaev--Roberts (GSR) change-point detection procedure, set up to react to a possible shift in the drift of standard Brownian motion monitored ``live''. Parameter $\mu$ present in the right-hand side of SDE~\eqref{eq:Rt_r-def} is the anticipated magnitude of the possible change in the drift. More concretely, equation~\eqref{eq:Rt_r-def} describes the dynamics of the GSR statistic $(R_{t}^r)_{t\ge0}$ in the pre-change regime, i.e., under the assumption that the drift $\mu$ has not yet ``kicked in'', so that the observed Brownian motion is still ``driftless''. Hence the stopping time $\mathcal{S}_A^r$ given by~\eqref{eq:T-GSR-def} is the GSR procedure's Run Length to {\em false alarm}: at time instance $\mathcal{S}_A^r$ the GSR procedure sounds a false alarm, i.e., falsely declares the Brownian motion under surveillance as having gained a drift of size $\mu\neq0$. The first moment of $\mathcal{S}_A^r$, i.e., $\EV[\mathcal{S}_A^r]$, is known in the change-point detection literature as the Average Run Length (ARL) to false alarm, and it is a popular metric of the ``cost'' of triggering a false alarm. Obviously $\EV[\mathcal{S}_A^r]$ increases with $A>0$, and, in particular, letting $A$ explode is the same as letting $\EV[\mathcal{S}_A^r]$ explode, and vice versa.

The GSR procedure was proposed by~\cite{Moustakides+etal:SS11} as a headstarted (hence, more general) version of the classical quasi-Bayesian Shiryaev--Roberts (SR) procedure that emerged from the independent work of Shiryaev~\citeyearpar{Shiryaev:SMD61,Shiryaev:TPA63} and that of Roberts~\citeyearpar{Roberts:T66}. The interest in the GSR procedure (and its variations) is due to its recently discovered strong optimality properties. See, e.g.,~\cite{Burnaev:ARSAIM2009},~\cite{Feinberg+Shiryaev:SD2006},~\cite{Burnaev+etal:TPA2009},~\cite{Pollak+Tartakovsky:SS2009},~\cite{Polunchenko+Tartakovsky:AS10},~\cite{Tartakovsky+Polunchenko:IWAP10},~\cite{Vexler+Gurevich:MESA2011}, and~\cite{Tartakovsky+etal:TPA2012}.

We are now in a position to describe the specific contribution of this work. It is shown in the sequel that a suitably standardized version of the stopping time $\mathcal{S}_A^r$ given by~\eqref{eq:T-GSR-def} is asymptotically, as $A\to+\infty$, exponentially distributed with unit mean, for any headstart $R_0^r\triangleq r\ge0$. Put another way, the GSR procedure's Run Length to false alarm, properly scaled, is asymptotically, as the ARL to false alarm level explodes (i.e., as $\EV[\mathcal{S}_A^r]\to+\infty$), unit-mean exponential. More specifically, it is shown in the sequel that, as GSR procedure's ARL to false alarm level gets large, the moment generating function (mgf) or the Laplace transform of a properly scaled version of the GSR stopping time $\mathcal{S}_A^r$ converges to that of the unit-mean exponential distribution. This implies convergence in distribution. The proof is explicit in that the mgf is first found analytically and in a closed form, and then the desired limit is shown directly to evaluate to the mgf of the unit-mean exponential distribution. It is also of note that the unit-drift assumption imposed on $(R_{t}^r)_{t\ge0}$ is essential, for it makes $(R_{t}^r)_{t\ge0}$ a (positive) recurrent process with all the ensuing consequences which ultimately ``add up'' to the desired asymptotic exponentiality of $\mathcal{S}_A^r$.

The discrete-time analogue of our result has been previously established---in an entirely different fashion---by Pollak and Tartakovsky~\citeyearpar{Pollak+Tartakovsky:TPA2009}; see also~\cite{Tartakovsky+etal:IWAP2008} and~\cite{Yakir:AS1998,Yakir:AS1995}. As a matter of fact, Pollak and Tartakovsky~\citeyearpar{Pollak+Tartakovsky:TPA2009} proved the result not only for the GSR procedure, but for an entire class of Markov stopping times, which includes the GSR procedure as well as Page's~\citeyearpar{Page:B54} celebrated Cumulative Sum (CUSUM) ``inspection'' scheme. More importantly, Pollak and Tartakovsky~\citeyearpar{Pollak+Tartakovsky:TPA2009} also illustrated the importance of the result in the context of sequential change-point detection. Specifically, they argued that if the stopping time of a change-point detection procedure is asymptotically exponential under the no-change hypothesis, it is reasonable to expect it to be approximately exponentially distributed (under the no-change hypothesis) whenever the ARL to false alarm is large. Consequently, since the exponential distribution is fully characterized but its mean alone, the ARL to false alarm can be seen as indeed being an exhaustive metric of the false alarm risk. See, e.g.,~\cite{Tartakovsky:SA08-discussion} for a more detailed discussion of this issue. Moreover, Pollak and Tartakovsky~\citeyearpar{Pollak+Tartakovsky:TPA2009} also argued that the asymptotic exponentiality (in the pre-change regime) can be used for the evaluation of the change-point detection procedure's local false alarm probabilities. As pointed out by~\cite{Tartakovsky:IEEE-CDC05} these probabilities are of importance in a variety of applications. All these considerations obviously apply to the continuous-time setting considered in this work as well.

The rest of the paper is three sections. The first one, Section~\ref{sec:main-result}, is the paper's main section, for this is where we formally state and then prove our main result. The second one, Section~\ref{sec:numerics}, is where we offer a short numerical study to complement and confirm our theoretical contribution experimentally. The third one, Section~\ref{sec:conclusion}, is where we make a few concluding remarks and draw a line under the entire paper.

\section{The Main Result}
\label{sec:main-result}

We first formally introduce the main object of study of this work. Let
\begin{align}\label{eq:T-GSR-mgf-def}
M(\alpha;A,x)
&\triangleq
\EV\left\{e^{-\alpha\mathcal{S}_A^{r=x}}\right\},
\;\;
\alpha\ge0,\; x\in[0,A],\; A>0,
\end{align}
denote the mgf (Laplace transform) of the stopping time $\mathcal{S}_A^r$ given by~\eqref{eq:T-GSR-def}. We are interested in the asymptotic behavior of $M(\alpha;A,x)$ as $A\to+\infty$. To that end, an important fact about $M(\alpha;A,x)$ is that, for any $\alpha\ge0$, $A>0$ and $x\in[0,+\infty)$, it can actually be expressed analytically and in closed form through the spectral characteristics of the second-order differential operator
\begin{align}\label{eq:Doperator-def}
\mathscr{D}
&\triangleq
\dfrac{\mu^2}{2}\dfrac{\partial^2}{\partial x^2} x^2
-
\dfrac{\partial}{\partial x},
\end{align}
i.e., the infinitesimal generator of the GSR diffusion $(R_t^r)_{t\ge0}$ governed by the SDE~\eqref{eq:Rt_r-def}. More concretely, the operator $\mathscr{D}$ is restricted to the state space of $(R_t^r)_{t\ge0}$, i.e., the interval $[0,A]$, $A>0$, and the relevant spectral characteristics of $\mathscr{D}$ are the solutions $\lambda$ and $u(x,\lambda)$ of the Sturm--Liouville problem $\big[\mathscr{D}\circ u\big](x,\lambda)=\lambda\,u(x,\lambda)$ or explicitly
\begin{align}\label{eq:master-eqn}
\dfrac{\mu^2}{2}\dfrac{d^2}{dx^2}\big[x^2\,u(x,\lambda)\big]
-
\dfrac{d}{dx}\big[u(x,\lambda)\big]
&=
\lambda\,u(x,\lambda),
\;\;
x\in[0,A],
\end{align}
subject to the boundary conditions
\begin{align}\label{eq:bnd-conds}
\lim_{x\to0+}\left\{\dfrac{\mu^2}{2}\dfrac{\partial}{\partial x}\big[x^2\,u(x,\lambda)\big]-u(x,\lambda)\right\}
&=
0
\;\;
\text{and}
\;\;
u(A,\lambda)
=
0,
\end{align}
which, translated into classical Feller's~\citeyearpar{Feller:AM1952} boundary classification, cast $x=0$ as an {\em entrance} boundary for $(R_t^r)_{t\ge0}$, and $x=A$ as an absorbing boundary for $(R_t^r)_{t\ge0}$, i.e., the process is ``killed'' once it hits the right end of the interval $[0,A]$; in ``differential equations speak'', the former condition is a Neumann-type boundary condition, while the latter condition is a Dirichlet-type boundary condition. It is apparent that the spectrum $\{\lambda\}$ of the operator $\mathscr{D}$ is dependent on $A>0$, and from now on, wherever necessary, we shall emphasize this dependence via the notation $\{\lambda_{A}\}$. Equation~\eqref{eq:master-eqn} subject to the boundary conditions~\eqref{eq:bnd-conds} is a Sturm--Liouville problem, and it has recently received a renewed burst of attention in the literature on mathematical finance and quickest change-point detection. See, e.g.,~\cite{Linetsky:OR2004,Linetsky:HandbookChapter2007},~\cite{Collet+etal:Book2013}, and notably~\cite{Polunchenko:SA2016,Polunchenko:SA2017}. The work of Polunchenko~\citeyearpar{Polunchenko:SA2017} will be referenced repeatedly throughout the sequel, following, for convenience, Polunchenko's~\citeyearpar{Polunchenko:SA2017} original notation.

We now turn to the work of~\cite{Linetsky:HandbookChapter2007} and recall a general result from the interface between stochastic processes and Sturm--Liouville operator theory (theory of second-order self-adjoint differential operators); see also~\cite[Chapter~4,~Section~4.6]{Ito+McKean:Book1974} and~\cite[Chapter~II,~Section~1.10]{Borodin+Salminen:Book2002}. Let $(X_t)_{t\ge0}$ be a one-dimensional, time-homogeneous, regular Markov diffusion whose state space is some interval $(e_1,e_2)\subseteq\mathbb{R}$, where $-\infty\le e_1<e_2\le\infty$, and such that $X_0=x\in(e_1,e_2)$ is fixed. If $(X_t)_{t\ge0}$ is generated by the SDE $dX_t=a(X_t)dt+\sqrt{b(X_t)}\,dB_t$ where the diffusion coefficient $b(x)$ is continuous and strictly positive inside $(e_1,e_2)$ and the drift coefficient $a(x)$ is continuous on $(e_1,e_2)$, then the Laplace transform of the nonnegative random variable $\mathcal{T}_y^x\triangleq\inf\{t\ge0\colon X_t=y\}$ with $\inf\{\varnothing\}=+\infty$ is given by
\begin{align}\label{eq:general-mgf-answer}
\EV\left\{e^{-\alpha\mathcal{T}_y^x}\right\}
&=
\begin{cases}
\dfrac{\varphi(x;\alpha)}{\varphi(y;\alpha)},&\;\text{for $x\le y$;}\\[1em]
\dfrac{\psi(x;\alpha)}{\psi(y;\alpha)},&\;\text{for $y \le x$,}
\end{cases}
\end{align}
where $\alpha>0$, and $\varphi(x;\alpha)$ and $\psi(x;\alpha)$ are two fundamental solutions $v(x;\alpha)$ of the equation
\begin{align}\label{eq:general-eigfun-eqn}
\dfrac{1}{2} b(x)\dfrac{\partial^2}{\partial x^2}\big[v(x;\alpha)\big]+a(x)\dfrac{\partial}{\partial x}\big[v(x;\alpha)\big]
&=
\alpha\, v(x;\alpha),\;\; x\in(e_1,e_2),
\end{align}
subject to appropriate boundary conditions. Specifically, these fundamental solutions can be made unique (up to a multiplicative constant factor dependent on $\alpha$ but independent of $x$) by requiring that $\psi(x;\alpha)$ be an increasing function of $x$ subject to a boundary condition at $e_1$, while $\varphi(x;\alpha)$ be a decreasing function of $x$, subject a boundary condition at $e_2$.

To translate the above to our specific problem~\eqref{eq:master-eqn}--\eqref{eq:bnd-conds} it suffices to note that equation~\eqref{eq:master-eqn} can be easily converted to an equation of the form~\eqref{eq:general-eigfun-eqn} by means of the integrating factor method. As a matter of fact, for our operator $\mathscr{D}$ given by~\eqref{eq:Doperator-def}, it has already been established, e.g., by~\cite{Polunchenko+Sokolov:MCAP2016} and by~\cite{Polunchenko:SA2016}, that
\begin{align}\label{eq:eigfun-gen-form}
\psi(x,\lambda)
&=
\dfrac{\mu^2 x}{2}\,e^{\tfrac{1}{\mu^2 x}} M_{1,\tfrac{\xi(\lambda)}{2}}\left(\dfrac{2}{\mu^2 x}\right)
\;\;
\text{and}
\;\;
\varphi(x,\lambda)
=
\dfrac{\mu^2 x}{2}\,e^{\tfrac{1}{\mu^2 x}} W_{1,\tfrac{\xi(\lambda)}{2}}\left(\dfrac{2}{\mu^2 x}\right),
\end{align}
where
\begin{align}\label{eq:xi-def}
\xi
&\equiv
\xi(\lambda)
\triangleq
\sqrt{1+\dfrac{8}{\mu^2}\lambda}
\;\;\text{so that}\;\;
\lambda
\equiv
\lambda(\xi)
=
\dfrac{\mu^2}{8}(\xi^2-1),
\end{align}
and where $M_{a,b}(z)$ and $W_{a,b}(z)$ denote the so-called Whittaker $M$ and $W$ functions, respectively. The Whittaker functions are defined as the two fundamental solutions of the classical Whittaker~\citeyearpar{Whittaker:BAMS1904} equation
\begin{align*}
\dfrac{\partial^2}{\partial z^2}\,w(z)+\left\{-\dfrac{1}{4}+\dfrac{a}{z}+\dfrac{1/4-b^2}{z^2}\right\}w(z)
&=
0,
\end{align*}
where $w(z)$ is the unknown function of $z\in\mathbb{C}$, and $a,b\in\mathbb{C}$ are specified parameters; see, e.g.,~\cite[Chapter~I]{Buchholz:Book1969}. The Whittaker functions are typically considered in the cut plane $|\arg(z)\,|<\pi$ to ensure they are not multi-valued. For an extensive study of these functions and various properties thereof, see, e.g.,~\cite{Slater:Book1960} and~\cite{Buchholz:Book1969}.

At this point, in view of~\eqref{eq:general-mgf-answer} and~\eqref{eq:eigfun-gen-form}, we can conclude at once that
\begin{align}\label{eq:mgf-answer}
M(\alpha;A,x)
&=
\dfrac{\dfrac{\mu^2 x}{2}\,e^{\tfrac{1}{\mu^2 x}}W_{1,\tfrac{1}{2}\xi(\alpha)}\left(\dfrac{2}{\mu^2 x}\right)}{\dfrac{\mu^2 A}{2}\,e^{\tfrac{1}{\mu^2 A}}W_{1,\tfrac{1}{2}\xi(\alpha)}\left(\dfrac{2}{\mu^2 A}\right)},
\;\;
\alpha\ge0,\;x\in[0,A],\; A>0,
\end{align}
where $\xi\equiv\xi(\lambda)$ is as in~\eqref{eq:xi-def}. Parenthetically, we remark that, apparently, this result, though relatively simple to obtain, was not previously known to the change-point detection community. It is also of note that the Laplace transform~\eqref{eq:mgf-answer} {\em can} be inverted to yield the (pre-change) distribution of the GSR stopping time $\mathcal{S}_A^r$, and the inversion has already been performed by~\cite{Polunchenko:SA2017}.

To proceed, observe that $\mathcal{S}_A^r$, by definition~\eqref{eq:T-GSR-def}, almost surely explodes as $A\to+\infty$. Hence, it shouldn't come as a surprise that, for any fixed $\alpha\ge0$ and $x\in[0,+\infty)$, the limit of $M(\alpha;A,x)$ as $A\to+\infty$ is zero. Heuristically, this can be seen directly from the definition~\eqref{eq:T-GSR-mgf-def}. More formally, one can appeal to the small-argument asymptotic behavior of the Whittaker $W$ function
\begin{align*}
W_{a,b}(x)
&\sim
\dfrac{\Gamma(2b)}{\Gamma(b-a+1/2)}\,x^{-b+\tfrac{1}{2}}\,e^{-\tfrac{x}{2}}
\;\;\text{as}\;\; x\to0+,
\end{align*}
where here and onward $\Gamma(z)$ denotes the Gamma function (see, e.g.,~\citealt[Chapter~6]{Abramowitz+Stegun:Handbook1964}), to first get
\begin{align}\label{eq:Whit-func-lim-zero}
\dfrac{\mu^2 A}{2}\,e^{\tfrac{1}{\mu^2 A}}W_{1,\tfrac{1}{2}\xi(\alpha)}\left(\dfrac{2}{\mu^2 A}\right)
&\sim
\dfrac{\Gamma(\xi(\alpha))}{\Gamma(\xi(\alpha)/2-1/2)}\,\left(\dfrac{\mu^2 A}{2}\right)^{\tfrac{1}{2}\xi(\alpha)+\tfrac{1}{2}}
\;\;\text{as}\;\; A\to+\infty,
\end{align}
so that
\begin{align*}
\lim_{A\to+\infty}\left\{\dfrac{\mu^2 A}{2}\,e^{\tfrac{1}{\mu^2 A}}W_{1,\tfrac{1}{2}\xi(\alpha)}\left(\dfrac{2}{\mu^2 A}\right)\right\}
&=
+\infty,
\end{align*}
because $\xi(\alpha)\ge 1$ for $\alpha\ge0$, and then conclude from the formula~\eqref{eq:mgf-answer} for the mgf $M(\alpha;A,x)$ that the latter does, in fact, go to zero as $A\to+\infty$.

However, as previously noted by~\cite{Pollak+Tartakovsky:TPA2009}, there is a way to rescale $\mathcal{S}_A^r$ so as to get it to converge to a meaningful random variable as $A\to+\infty$; see also~\cite{Tartakovsky+etal:IWAP2008}. We now explain the idea.

Let
\begin{align*}
Q_A(x)
&\triangleq
\lim_{t\to+\infty}\Pr(R_t^r\le x|\mathcal{S}_A^r>t)
\;\;
\text{and}
\;\;
q_A(x)
\triangleq
\frac{d}{dx}Q_A(x),
\;\;
x\in[0,A],
\end{align*}
denote the GSR statistic's so-called quasi-stationary cumulative distribution function (cdf) and density, respectively. This time-invariant probability measure is independent of the GSR statistic's headstart $R_0^r\triangleq r\in[0,A]$, and its existence can be inferred, e.g., from the work of~\cite{Cattiaux+etal:AP2009}; see also~\cite[Section~7.8.2]{Collet+etal:Book2013}. Exact closed-form formulae for both $Q_A(x)$ and $q_A(x)$ have been recently obtained by~\cite{Polunchenko:SA2017}. If the GSR statistic $(R_t^r)_{t\ge0}$ is started off a random point sampled from its quasi-stationary distribution, i.e., if $R_0^r\triangleq r\propto Q_A(x)$, then the statistical characteristics of the GSR statistic will be time-invariant, until the statistic hits the threshold $A$. Since the probability of hitting $A$ will be time-invariant as well, the distribution of the GSR stopping time will be exponential. More formally, define $(R_t^Q)_{t\ge0}$ as the solution of the SDE $dR_t^Q=dt+\mu R_t^Q dB_t$ with $R_0^Q\propto Q_A(x)$, and let $\mathcal{S}_A^Q\triangleq\inf\{t\ge0\colon R_t^Q=A\}$ with $\inf\{\varnothing\}=+\infty$ and $A>0$. The stopping time $\mathcal{S}_A^Q$ is known in the quickest change-point detection literature as the randomized Shiryaev--Roberts--Pollak detection procedure, and it was originally proposed (for the discrete-time version of the problem) and first investigated by~\cite{Pollak:AS85}; it was also recently studied by~\cite{Burnaev+etal:TPA2009}. Specifically, since $(0\ge)\,\lambda_A\triangleq\log\Pr(R_t^Q\ge A|\mathcal{S}_A^Q>t)$ is level for all $t\ge0$, one can conclude that $\mathcal{S}_A^Q$ is exponentially distributed with parameter $-\lambda_A\,(>0)$, so that the product $-\lambda_A\mathcal{S}_A^Q$ is unit-mean exponential. As noted by~\cite{Pollak+Tartakovsky:TPA2009}, intuitively, the large-$A$ behavior of $\mathcal{S}_A^r$ for each fixed headstart is similar to that of $\mathcal{S}_A^Q$. Hence, it stands to reason that $-\lambda_A\mathcal{S}_A^r$ is approximately unit-mean exponential, whenever $A$ is large. Put another way, the right scaling factor for $\mathcal{S}_A^r$ is $-\lambda_A\triangleq-\log\Pr(R_t^Q\ge A|\mathcal{S}_A^Q>t)$.

The constant $\lambda_A\triangleq\log\Pr(R_t^Q\ge A|\mathcal{S}_A^Q>t)$ is the largest (nonpositive) eigenvalue of the operator $\mathscr{D}$ given by~\eqref{eq:Doperator-def}. For this kind of an operator it is known from the general Sturm--Liouville theory (see, e.g.,~\citealt{Fulton+etal:FIT-TR1999}) that its spectrum $\{\lambda\}$ is purely discrete, simple, located to the left of the origin (i.e., nonpositive), and is determined entirely by the Dirichlet condition~\eqref{eq:bnd-conds}, i.e., from the equation $u(A,\lambda)=0$ with $A>0$ fixed. More concretely, from~\eqref{eq:bnd-conds} and~\eqref{eq:eigfun-gen-form} it can be readily seen that $\lambda_A$ is the largest (nonpositive) solution of the equation
\begin{align}\label{eq:eigval-eqn}
W_{1,\tfrac{1}{2}\xi(\lambda_A)}\left(\dfrac{2}{\mu^2 A}\right)
&=
0,
\end{align}
where $A>0$ is fixed and $\xi(\lambda)$ is as in~\eqref{eq:xi-def}. This equation was previously analyzed by~\cite{Polunchenko:SA2016,Polunchenko:SA2017}, who, in particular, obtained order-one, order-two, and order-three asymptotic ``large-$A$'' approximations to $\lambda_A$. As an aside, we note that due to the discrete and simple nature of the spectrum of the operator $\mathscr{D}$ the range of values of $\alpha$ in the above formula~\eqref{eq:mgf-answer} for the mgf $M(\alpha;A,x)$ can be extended from $\alpha\in[0,+\infty)$ to $\alpha\in(\lambda_A,+\infty)$ where $\lambda_A\le 0$ is largest (nonpositive) eigenvalue of $\mathscr{D}$.

The main contribution of this work can now be succinctly put as follows.
\begin{theorem}\label{thm:main-result}
$\lim_{A\to+\infty} M(-\alpha\lambda_A;A,x)=1/(1+\alpha)$ for any fixed $\alpha\in(-1,+\infty)$ and $x\in[0,+\infty)$; recall that $\lambda_A$ here is the largest (nonpositive) solution of equation~\eqref{eq:eigval-eqn}.
\end{theorem}

The plan for the remainder of this section is to prove this theorem. To that end, in view of~\eqref{eq:master-eqn}, the problem essentially is to show that
\begin{align}\label{eq:mgf-lim-claim}
\lim_{A\to+\infty}\left\{\dfrac{\dfrac{\mu^2 x}{2}\,e^{\tfrac{1}{\mu^2 x}}W_{1,\tfrac{1}{2}\xi(-\alpha\lambda_A)}\left(\dfrac{2}{\mu^2 x}\right)}{\dfrac{\mu^2 A}{2}\,e^{\tfrac{1}{\mu^2 A}}W_{1,\tfrac{1}{2}\xi(-\alpha\lambda_A)}\left(\dfrac{2}{\mu^2 A}\right)}\right\}
&=
\dfrac{1}{1+\alpha}
\;\;
\text{for any fixed}
\;\;
\alpha\in(-1,+\infty)
\;\;
\text{and}
\;\;
x\in[0,+\infty),
\end{align}
where $\xi=\xi(\lambda)$ is as in~\eqref{eq:xi-def} and $\lambda_A$ is the largest (nonpositive) solution of equation~\eqref{eq:eigval-eqn}.

The above limit can be evaluated by treating the numerator and the denominator separately. The key observation for either part is that $\lambda_A\nearrow 0$ as $A\to+\infty$, i.e., $\lambda_A$ is a monotonically increasing function of $A>0$, converging to 0 from below as $A\to+\infty$; see~\cite{Polunchenko:SA2017} for a proof. An immediate implication of this circumstance is that since $\lim_{A\to+\infty}\lambda_A=0$, then from~\eqref{eq:xi-def} we also have $\lim_{A\to+\infty}\xi(\lambda_A)=1$, and therefore
\begin{align}\label{eq:mgf-lim-numerator}
\lim_{A\to+\infty}\left\{\dfrac{\mu^2 x}{2}\,e^{\tfrac{1}{\mu^2 x}} W_{1,\tfrac{1}{2}\xi(-\alpha\lambda_A)}\left(\dfrac{2}{\mu^2 x}\right)\right\}
&=
1,
\end{align}
because $W_{1,\tfrac{1}{2}}(z)=z\,e^{-\tfrac{z}{2}}$ which is a special case of~\cite[Identity~(28a),~p.~23]{Buchholz:Book1969} asserting that $W_{a,a-\tfrac{1}{2}}(z)=z^{a}e^{-\tfrac{z}{2}}$. Hence, the numerator of the fraction under the limit~\eqref{eq:mgf-lim-claim} goes to unity as $A\to+\infty$.

It remains to take care of the denominator of the fraction under the limit~\eqref{eq:mgf-lim-claim}, i.e., to show that
\begin{align}\label{eq:mgf-lim-denominator}
\lim_{A\to+\infty}\left\{\dfrac{\mu^2 A}{2}\,e^{\tfrac{1}{\mu^2 A}}W_{1,\tfrac{1}{2}\xi(-\alpha\lambda_A)}\left(\dfrac{2}{\mu^2 A}\right)\right\}
&=
1+\alpha
\;\;
\text{for all}
\;\;
\alpha\in(-1,+\infty),
\end{align}
which is a more delicate problem. Specifically, the problem is that not only the argument of the Whittaker $W$ function is dependent on $A$, but also its second index $\xi(\lambda_A)/2$ which goes to $1/2$ as $A\to+\infty$ because $\lim_{A\to+\infty}\lambda_A=0$. As a result, the above small-argument asymptotics~\eqref{eq:Whit-func-lim-zero} of the Whittaker $W$ function is not ``fine'' enough and needs to be improved.

To that end, let us again turn to the work of~\cite{Polunchenko:SA2017} where the function $f(\lambda)\triangleq W_{1,\tfrac{1}{2}\xi(\lambda)}(z)$ was expanded into a Taylor series with respect to $\lambda$ around zero up to the third order for any fixed $z\ge0$; it is noteworthy that $W_{a,b}(z)$ is an entire function of $b\in\mathbb{C}$ for any fixed $a\in\mathbb{R}$ and $z>0$. The expansion involves the following two special functions:
\begin{itemize}
    \item The exponential integral
\begin{align}\label{eq:E1-func-def}
\E1(x)
&\triangleq
\int_{x}^\infty\dfrac{e^{-y}}{y}\,dy,\; x>0;
\end{align}
see, e.g.,~\cite[Chapter~5]{Abramowitz+Stegun:Handbook1964}; and
    \item Meijer's~\citeyearpar{Meijer:NAW1936} celebrated $G$-function defined as the Mellin-Barnes integral
\begin{align*}
\MeijerG*{m}{n}{p}{q}{a_1,\ldots,a_p}{b_1,\ldots,b_q}{z}
&\triangleq
\dfrac{1}{2\pi\imath}\int_{\mathcal{C}}\dfrac{\prod_{k=1}^m \Gamma(b_k-s) \prod_{j=1}^n \Gamma(1-a_j+s)}{\prod_{k=m+1}^q \Gamma(1-b_k+s) \prod_{j=n+1}^p \Gamma(a_j-s)}\,z^s ds,
\end{align*}
where $\imath$ denotes the imaginary unit, i.e., $\imath\triangleq\sqrt{-1}$, the integers $m$, $n$, $p$, and $q$ are such that $0\le m\le q$ and $0\le n\le p$, and the contour of integration $\mathcal{C}$ is closed in an appropriate way to ensure the convergence of the integral. It is also required that no difference $a_j-b_k$ be an integer. The $G$-function is a very general function, and includes, as special cases, not only all elementary functions, but a number of special functions as well. An extensive list of special cases of the Meijer $G$-function can be found, e.g., in the classical special functions handbook of~\cite{Prudnikov+etal:Book1990}, which also includes a summary of the function's basic properties. We will need the following particular case of the Meijer $G$-function:
\begin{align}\label{eq:MeijerG-func-spcl-case}
\MeijerG*{3}{1}{2}{3}{0,1}{0,0,0}{x}
&=
\int_{x}^{+\infty}e^{y}\E1(y)\,\dfrac{dy}{y},\; x>0,
\end{align}
where $\E1(z)$ is the exponential integral defined in~\eqref{eq:E1-func-def}; see~\cite[Appendix~A]{Polunchenko:SA2017} for a proof.
\end{itemize}

We are now in a position present the third-order Taylor expansion obtained by~\cite{Polunchenko:SA2017} for the function $f(\lambda)\triangleq W_{1,\tfrac{1}{2}\xi(\lambda)}(z)$ with $z\ge0$ fixed.
\begin{theorem}[\citealt{Polunchenko:SA2017}]\label{thm:Whit-func-Taylor-expn}
For any $x\ge0$ it holds true that
\begin{align*}
\begin{split}
W_{1,\tfrac{1}{2}\xi(\lambda)}&\left(\dfrac{2}{\mu^2 x}\right)
=
\dfrac{2}{\mu^2}\,e^{-\tfrac{1}{\mu^2 x}}\Biggl\{\dfrac{1}{x}+\lambda+\dfrac{2}{\mu^2}L\left(\dfrac{2}{\mu^2 x}\right)\lambda^2+\\
&\quad+
\left(\dfrac{2}{\mu^2}\right)^2\left[\MeijerG*{3}{1}{2}{3}{0,1}{0,0,0}{\dfrac{2}{\mu^2 x}}-2L\left(\dfrac{2}{\mu^2 x}\right)\right]\lambda^3\Biggr\}
+\mathcal{O}(\lambda^4),
\end{split}
\end{align*}
where $\xi(\lambda)$ is as in~\eqref{eq:xi-def}, and
\begin{align}\label{eq:LwrBnd-def}
L(x)
&\triangleq
e^{x}\E1(x)-1+x\,\MeijerG*{3}{1}{2}{3}{0,1}{0,0,0}{x},
\end{align}
with $\E1(x)$ and $\MeijerG*{3}{1}{2}{3}{0,1}{0,0,0}{x}$ given by~\eqref{eq:E1-func-def} and by~\eqref{eq:MeijerG-func-spcl-case}, respectively.
\end{theorem}

This theorem readily gives the expansion
\begin{align}\label{eq:Whit-func-expn}
\begin{split}
\dfrac{\mu^2 A}{2}\,e^{\tfrac{1}{\mu^2 A}}W_{1,\tfrac{1}{2}\xi(-\alpha\lambda_A)}&\left(\dfrac{2}{\mu^2 A}\right)
=
1-A(\alpha\lambda_A)+A\dfrac{2}{\mu^2}L\left(\dfrac{2}{\mu^2 A}\right)(\alpha\lambda_A)^2+\\
&\quad
-A\left(\dfrac{2}{\mu^2}\right)^2\left[\MeijerG*{3}{1}{2}{3}{0,1}{0,0,0}{\dfrac{2}{\mu^2 A}}-2L\left(\dfrac{2}{\mu^2 A}\right)\right](\alpha\lambda_A)^3
+A \alpha^4\mathcal{O}(\lambda_A^4),
\end{split}
\end{align}
which, as we shall see shortly, is more ``fine'' than necessary to pass $A\to+\infty$ and prove~\eqref{eq:mgf-lim-denominator}. To do so, we first recall yet another result of~\cite{Polunchenko:SA2017}, viz. the double inequality
\begin{align*}
-\dfrac{1}{A}
&\le
\lambda_A
\le
-\dfrac{1}{A}-\dfrac{1-\sqrt{4\mu^2 A+1}}{2\mu^2 A^2}\;(<0),
\;
\text{for any}
\;
A>0,
\end{align*}
where $\mu\neq0$ is the parameter of the SDE~\eqref{eq:Rt_r-def}. Hence $\lambda_A=-1/A+\mathcal{O}(A^{-3/2})$ so that
\begin{align}\label{eq:eigval-limits}
\lim_{A\to+\infty}[A(-\lambda_A)]
&=
1
\;\;
\text{but}
\;\;
\lim_{A\to+\infty}[A(-\lambda_A)^{1+s}]
=
0
\;\;
\text{for $s>0$}.
\end{align}

The only issue is that the functions $L(x)$ and $\MeijerG*{3}{1}{2}{3}{0,1}{0,0,0}{x}$ given, respectively, by~\eqref{eq:LwrBnd-def} and~\eqref{eq:MeijerG-func-spcl-case}, both go to infinity as $x$ goes to zero. However, in view of~\cite[Inequality~5.1.20,~p.~229]{Abramowitz+Stegun:Handbook1964} which states that
\begin{align*}
\dfrac{1}{2}\log\left(1+\dfrac{2}{x}\right)
&<
e^x\E1(x)
<
\log\left(1+\dfrac{1}{x}\right)
\;\;
\text{for $x>0$},
\end{align*}
from~\eqref{eq:E1-func-def},~\eqref{eq:LwrBnd-def} and~\eqref{eq:MeijerG-func-spcl-case} it can be seen that
\begin{align}\label{eq:LG-funcs-limits}
\lim_{x\to+\infty}\left\{\dfrac{1}{x}L\left(\dfrac{1}{x}\right)\right\}
&=
0
\;\;\text{and}\;\;
\lim_{x\to+\infty}\left\{\dfrac{1}{x}\MeijerG*{3}{1}{2}{3}{0,1}{0,0,0}{\dfrac{1}{x}}\right\}
=
0,
\end{align}
i.e., the functions $L(1/x)$ and $\MeijerG*{3}{1}{2}{3}{0,1}{0,0,0}{1/x}$ both go to infinity as $x\to+\infty$ {\em slower} than $1/x$ goes to zero as $x\to+\infty$.

At this point Theorem~\ref{thm:main-result}, which is our main result, is straightforward to prove: it is merely a matter of using~\eqref{eq:eigval-limits} and~\eqref{eq:LG-funcs-limits} in~\eqref{eq:Whit-func-expn} to obtain~\eqref{eq:mgf-lim-denominator}, and then combining it with~\eqref{eq:mgf-lim-numerator} to get~\eqref{eq:mgf-lim-claim}, which in view of~\eqref{eq:mgf-answer} is precisely the desired result.

To draw a line under this section, we remark that the formula~\eqref{eq:mgf-answer} for the mgf $M(\alpha;A,x)$ of the GSR stopping time $\mathcal{S}_A^r$ can also be put to a more classical use, viz. to compute $\EV[(\mathcal{S}_A^r)^n]$ for $n\ge1$, i.e., to determine the actual moments of the GSR stopping time (under the no-change hypothesis). Specifically, since from definition~\eqref{eq:T-GSR-mgf-def} it is evident that
\begin{align*}
\EV[(\mathcal{S}_A^{r=x})^n]
&=
(-1)^n\left.\left[\dfrac{\partial^n}{\partial\alpha^n}M(\alpha;A,x)\right]\right|_{\alpha=0}, \;\; n\ge1,
\end{align*}
and because formula~\eqref{eq:mgf-answer} expresses $M(\alpha;A,x)$ explicitly as a quotient of two Whittaker $W$ functions, getting the $n$-th moment of the GSR stopping time essentially comes down to finding the derivatives, up through the $n$-th order inclusive, of the Whittaker $W$ function $W_{a,b}(z)$ with respect to the second index $b$. More concretely, from~\eqref{eq:mgf-answer} it is direct to see that the required derivatives are of the following form:
\begin{align*}
&\left.\left\{\dfrac{\partial^n}{\partial\alpha^n}\left[\dfrac{\mu^2 x}{2}\,e^{\tfrac{1}{\mu^2 x}} W_{1,\tfrac{1}{2}\xi(\alpha)}\left(\dfrac{2}{\mu^2 x}\right)\right]\right\}\right|_{\alpha=0},
\;\;
n\ge1,
\end{align*}
where $x\in[0,A]$, $A>0$, and $\xi(\lambda)$ is as in~\eqref{eq:xi-def}. Since the first three (for $n=1$, $2$ and $3$) of these derivatives are essentially given by Theorem~\ref{thm:Whit-func-Taylor-expn} due to~\cite{Polunchenko:SA2017}, computing $\EV[\mathcal{S}_A^r]$, $\EV[(\mathcal{S}_A^r)^2]$, and $\EV[(\mathcal{S}_A^r)^3]$, i.e., the first three moments of the GSR stopping time, is a matter of elementary algebra. The answer is:
\begin{align*}
\EV[\mathcal{S}_A^r]
&=
A-r,
\;\;
\EV[(\mathcal{S}_A^r)^2]
=
\dfrac{4}{\mu^2}\left[rL\left(\dfrac{2}{\mu^2 r}\right)-AL\left(\dfrac{2}{\mu^2 A}\right)\right]-2A(r-A),
\end{align*}
and
\begin{align*}
\begin{split}
\EV[(\mathcal{S}_A^r)^3]
&=
-6\left(\dfrac{2}{\mu^2}\right)^2\left\{\left[r\,\MeijerG*{3}{1}{2}{3}{0,1}{0,0,0}{\dfrac{2}{\mu^2 r}}-2rL\left(\dfrac{2}{\mu^2 r}\right)\right]-\left[A\,\MeijerG*{3}{1}{2}{3}{0,1}{0,0,0}{\dfrac{2}{\mu^2 A}}-2AL\left(\dfrac{2}{\mu^2 A}\right)\right]\right\}+\\
&
\qquad\qquad
+6A\dfrac{2}{\mu^2}\left[rL\left(\dfrac{2}{\mu^2 r}\right)-2AL\left(\dfrac{2}{\mu^2 A}\right)+rL\left(\dfrac{2}{\mu^2 A}\right)\right]
+6A^2(r-A).
\end{split}
\end{align*}
where $r\in[0,A]$, $A>0$, and the functions $L(x)$ and $\MeijerG*{3}{1}{2}{3}{0,1}{0,0,0}{x}$ are given, respectively, by~\eqref{eq:LwrBnd-def} and~\eqref{eq:MeijerG-func-spcl-case}. The first moment formula $\EV[\mathcal{S}_A^r]=A-r$ is well-known in quickest change-point detection, and was obtained---in an entirely different fashion---by~\cite{Shiryaev:SMD61,Shiryaev:TPA63} and many others. However, the second and third moment formulae appear to be new results. The fourth and higher moments can be found in a similar fashion, but since the formulae are far more cumbersome, they will be presented elsewhere.


\section{Numerical Results}
\label{sec:numerics}

To get a better sense as to how fast, as $A\to+\infty$, the random variable $-\lambda_A\mathcal{S}_A^r$ becomes unit-mean exponentially-distributed, we now offer a short numerical study where we assess the proximity of $\Pr(-\lambda_A\mathcal{S}_A^r\ge t)$ to $e^{-t}$ for various values of $A>0$, $r\in[0,A]$, $\mu\neq0$, and $t\ge0$. Specifically, since from Theorem~\ref{thm:main-result} we can deduce that $\lim_{A\to+\infty}\Pr(-\lambda_A\mathcal{S}_A^r\ge t)=e^{-t}$ for any $t\ge0$, or equivalently that $\lim_{A\to+\infty}\log\Pr(-\lambda_A\mathcal{S}_A^r\ge t)=-t$ for any $t\ge0$, it is reasonable to expect the function $f(t)\triangleq \log\Pr(-\lambda_A\mathcal{S}_A^r\ge t)$ to be close to $-t$ for any $t\ge0$, provided, however, that $A>0$ is sufficiently large. It is the proximity of $f(t)\triangleq \log\Pr(-\lambda_A\mathcal{S}_A^r\ge t)$ to the line $-t$ across a range of values of $t\ge0$ that we shall use to judge how close the distribution of $-\lambda_A\mathcal{S}_A^r$ is to unit-mean exponential. The evaluation of $f(t)\triangleq \log\Pr(-\lambda_A\mathcal{S}_A^r\ge t)$ as a function of $t$ for any $A>0$, $r\in[0,A]$, and $\mu\neq 0$ is not a problem at all, because the survival function $\Pr(\mathcal{S}_A^r\ge t)$ of the GSR stopping time $\mathcal{S}_A^r$ given by~\eqref{eq:T-GSR-def} was recently found analytically and in a closed-form by~\cite{Polunchenko:SA2016} who also developed a Mathematica script to evaluate $\Pr(\mathcal{S}_A^r\ge t)$ and $\lambda_A$ each to within hundreds of decimal places of accuracy and for any $A>0$, $r\in[0,A]$, and $\mu\neq 0$.

To get started, let us first set $A=100$, which, in practice, would be considered low, so that the asymptotic exponentiality might not be quite in effect yet. Figures~\ref{fig:logP-vs-t-A100-r0} show the obtained results for $t\in[0,10]$, $A=100$, $r=0$, and $\mu=\{1/2,1,3/2\}$. Specifically, Figure~\ref{fig:logP-vs-t-A100-r0:logP} shows $\log\Pr(-\lambda_A\mathcal{S}_A^r\ge t)$ as a function of $t\in[0,10]$, while Figure~\ref{fig:logP-vs-t-A100-r0:abs-err} shows the corresponding absolute error $\abs{-\log\Pr(-\lambda_A\mathcal{S}_A^r\ge t)-t}$. For convenience, Figure~\ref{fig:logP-vs-t-A100-r0:logP} also includes the line $-t$ which $\log\Pr(-\lambda_A\mathcal{S}_A^r\ge t)$ is to converge to as $A\to+\infty$. An eye examination of these figures suggests that the distribution of $-\lambda_A\mathcal{S}_A^r$ nearly unit-mean exponential, even though $A$ is as low as $100$. The agreement with the asymptotic exponential distribution is even better when $A$ is larger.
\begin{figure}[!htbp]
    \centering
    \begin{subfigure}{0.48\textwidth}
        \centering
        \includegraphics[width=\linewidth]{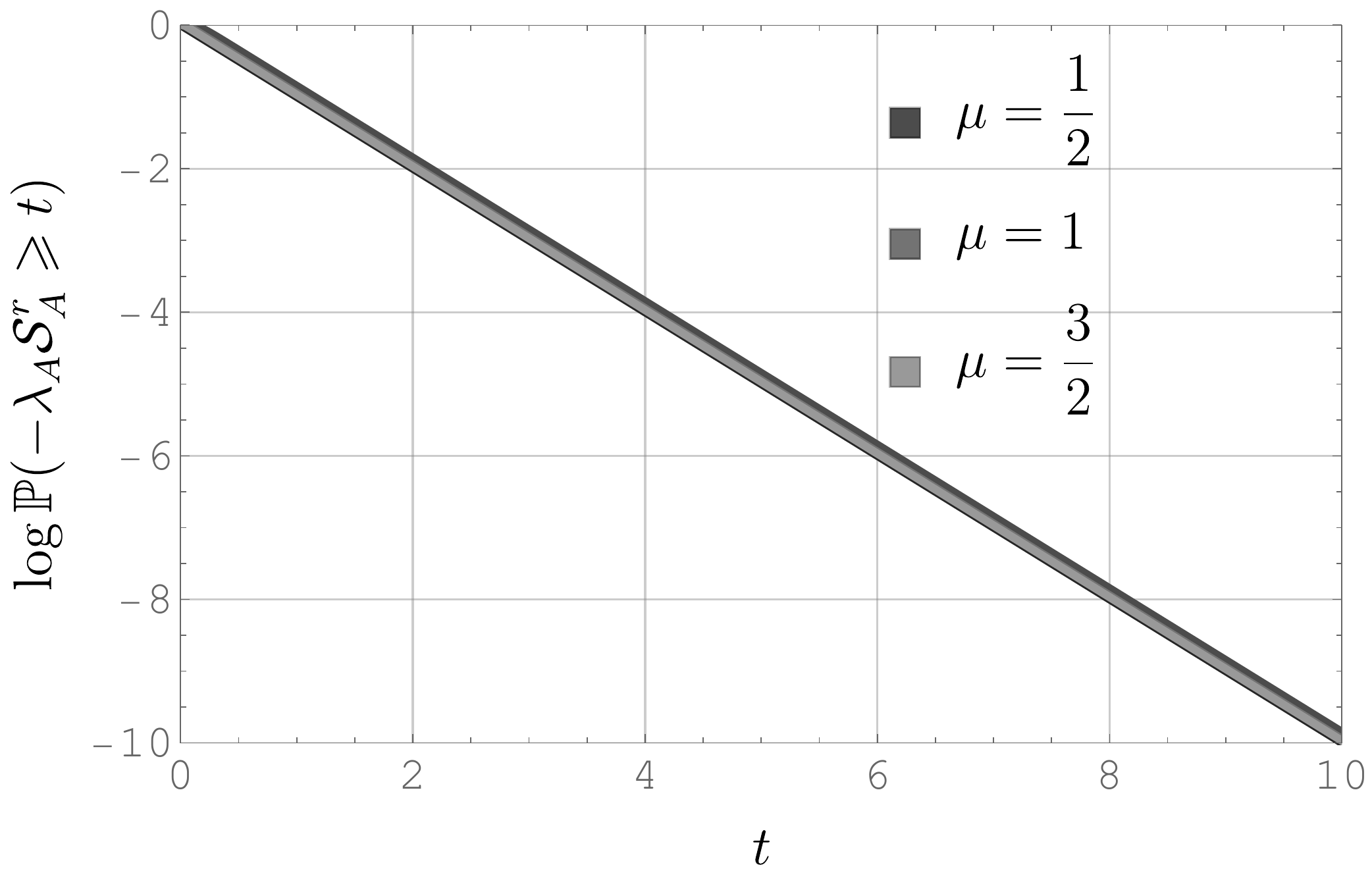}
        \caption{}
        \label{fig:logP-vs-t-A100-r0:logP}
    \end{subfigure}
    \hspace*{\fill}
    \begin{subfigure}{0.48\textwidth}
        \centering
        \includegraphics[width=\linewidth]{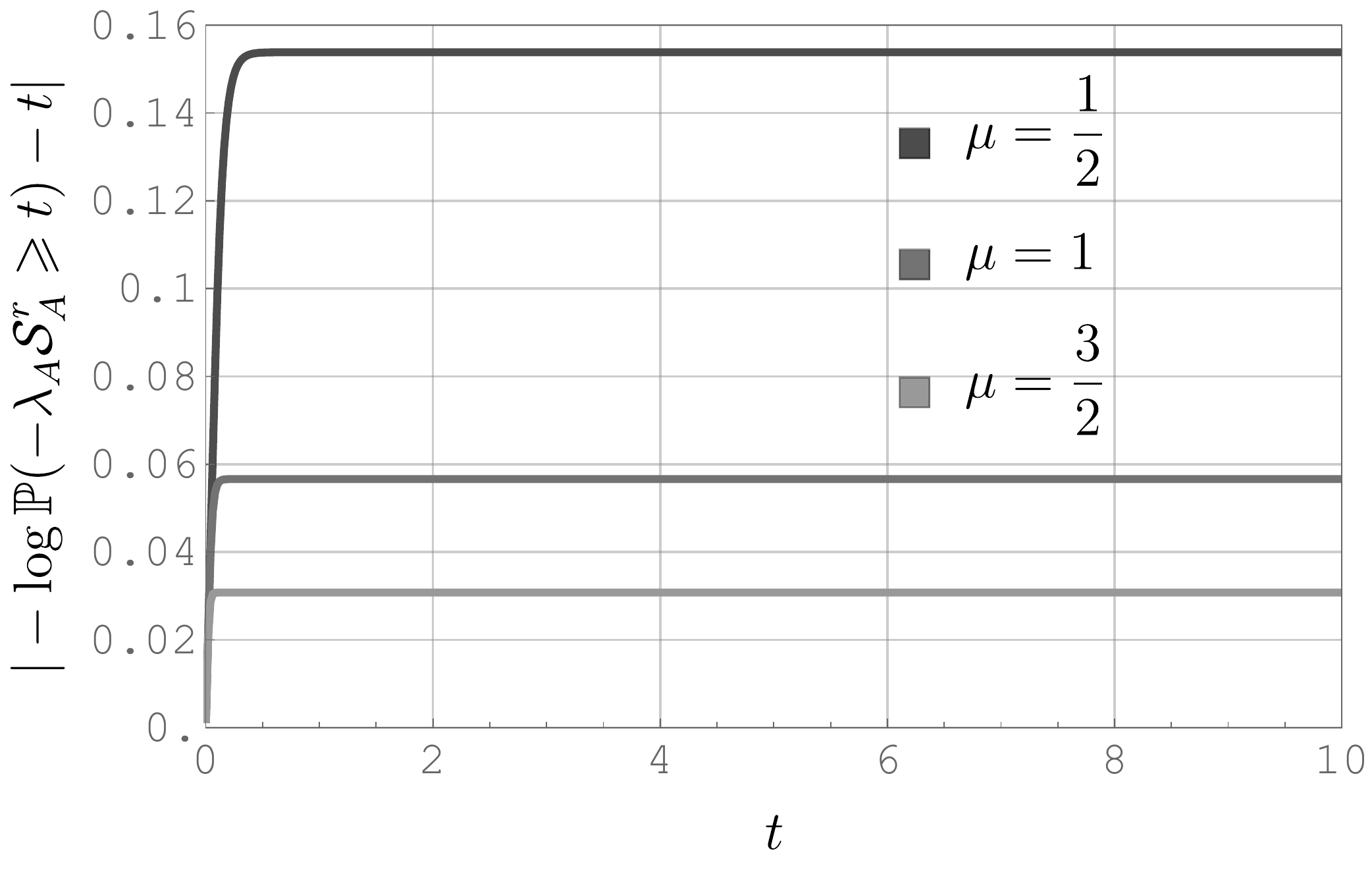}
        \caption{}
        \label{fig:logP-vs-t-A100-r0:abs-err}
    \end{subfigure}
    \caption{$\log\Pr(-\lambda_A\mathcal{S}_A^r\ge t)$ and $\abs{-\log\Pr(-\lambda_A\mathcal{S}_A^r\ge t)-t}$ as functions of $t$ for $t\in[0,10]$, $A=100$, $r=0$, and $\mu=\{1/2,1,3/2\}$.}
    \label{fig:logP-vs-t-A100-r0}
\end{figure}

Let us now keep $A$ at $100$ but increase the GSR statistic's headstart $R_0^r\triangleq r$ to $r=50$. Since $A$ is only $100$, setting $r$ to half that is bringing $(R_t^r)_{t\ge0}$ much closer to $A$, thereby aiding the former to hit the latter sooner (on average). Put another way, increasing the headstart $r$ is, in some sense, akin to lowering the threshold $A>0$. As a result, the asymptotic exponentiality might not ``kick in'' as fast. This is exactly what we see in Figures~\ref{fig:logP-vs-t-A100-r50:logP} and~\ref{fig:logP-vs-t-A100-r50:abs-err} which show the obtained results for $t\in[0,10]$, $A=100$, $r=50$, and $\mu=\{1/2,1,3/2\}$. Again, Figure~\ref{fig:logP-vs-t-A100-r50:logP} also includes the line $-t$, but this time around the deviation of $\log\Pr(-\lambda_A\mathcal{S}_A^r\ge t)$ from $-t$ is noticeable with a naked eye. This is evidence that the asymptotic exponentiality isn't quite there yet, and it is a direct consequence of the higher headstart value $r$.
\begin{figure}[!htbp]
    \centering
    \begin{subfigure}{0.48\textwidth}
        \centering
        \includegraphics[width=\linewidth]{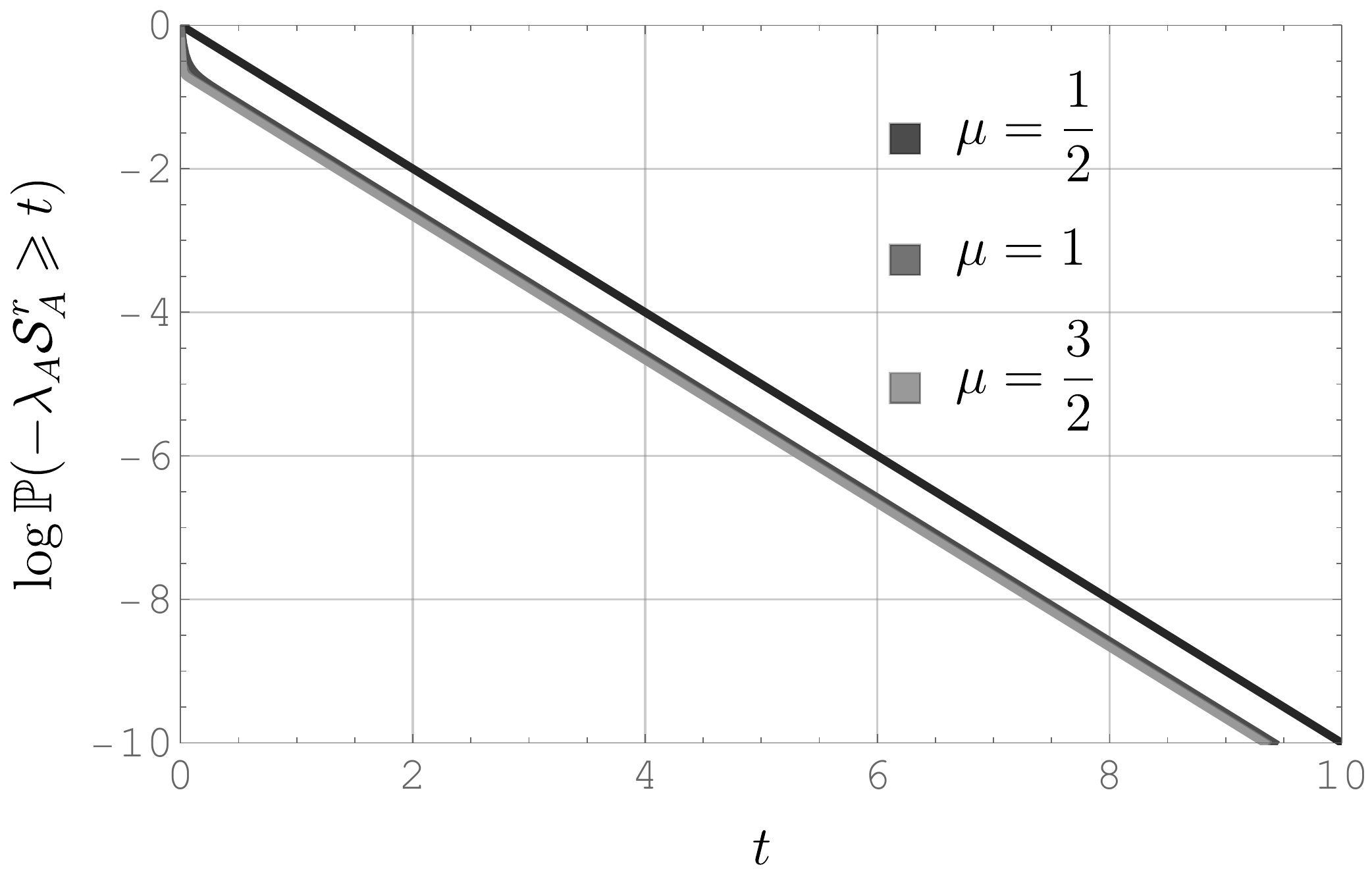}
        \caption{}
        \label{fig:logP-vs-t-A100-r50:logP}
    \end{subfigure}
    \hspace*{\fill}
    \begin{subfigure}{0.48\textwidth}
        \centering
        \includegraphics[width=\linewidth]{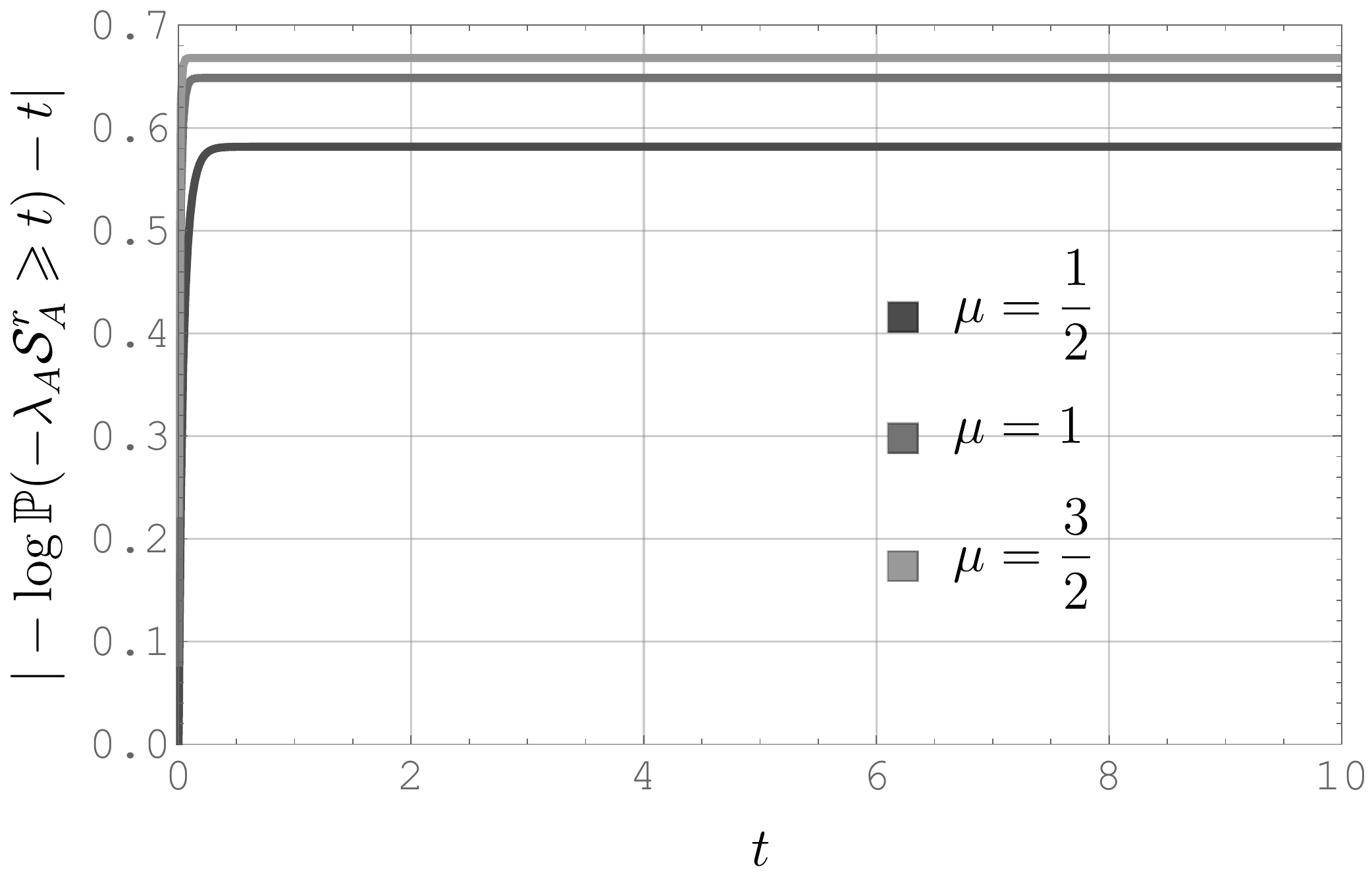}
        \caption{}
        \label{fig:logP-vs-t-A100-r50:abs-err}
    \end{subfigure}
    \caption{$\log\Pr(-\lambda_A\mathcal{S}_A^r\ge t)$ and $\abs{-\log\Pr(-\lambda_A\mathcal{S}_A^r\ge t)-t}$ as functions of $t$ for $t\in[0,10]$, $A=100$, $r=50$, and $\mu=\{1/2,1,3/2\}$.}
    \label{fig:logP-vs-t-A100-r50}
\end{figure}

However, if we keep $r$ at $50$ but increase $A$ to $500$, the distribution will get better aligned with the limiting exponential distribution, as can be seen from Figures~\ref{fig:logP-vs-t-A500-r50} which show the results for $t\in[0,10]$, $A=500$, $r=50$, and $\mu=\{1/2,1,3/2\}$. Looking at Figures~\ref{fig:logP-vs-t-A500-r50:logP} and~\ref{fig:logP-vs-t-A500-r50:abs-err} we see that $-\lambda_A\mathcal{S}_A^r$ is fairly close to being a unit-mean exponential random variable.
\begin{figure}[!htbp]
    \centering
    \begin{subfigure}{0.48\textwidth}
        \centering
        \includegraphics[width=\linewidth]{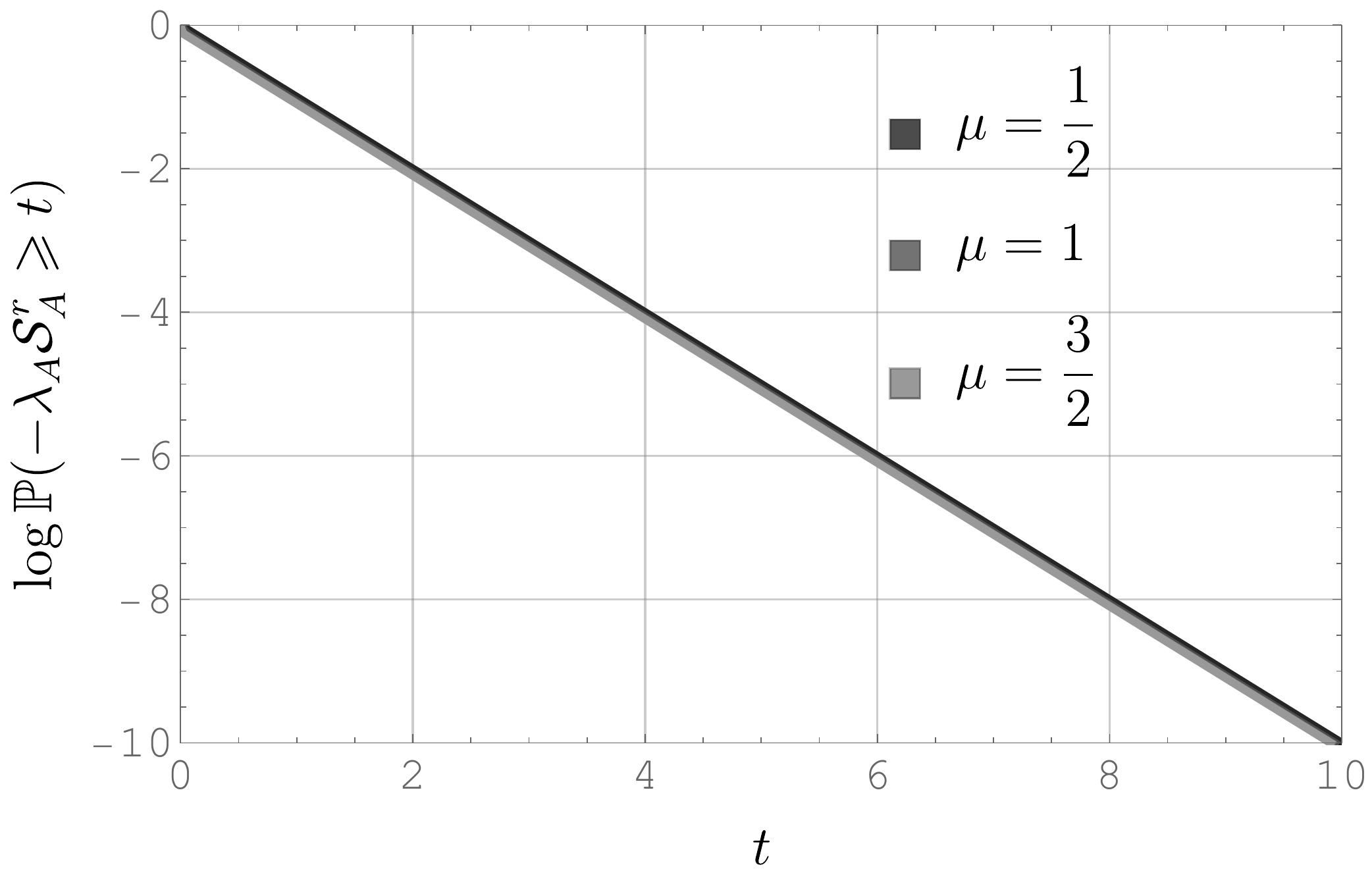}
        \caption{}
        \label{fig:logP-vs-t-A500-r50:logP}
    \end{subfigure}
    \hspace*{\fill}
    \begin{subfigure}{0.48\textwidth}
        \centering
        \includegraphics[width=\linewidth]{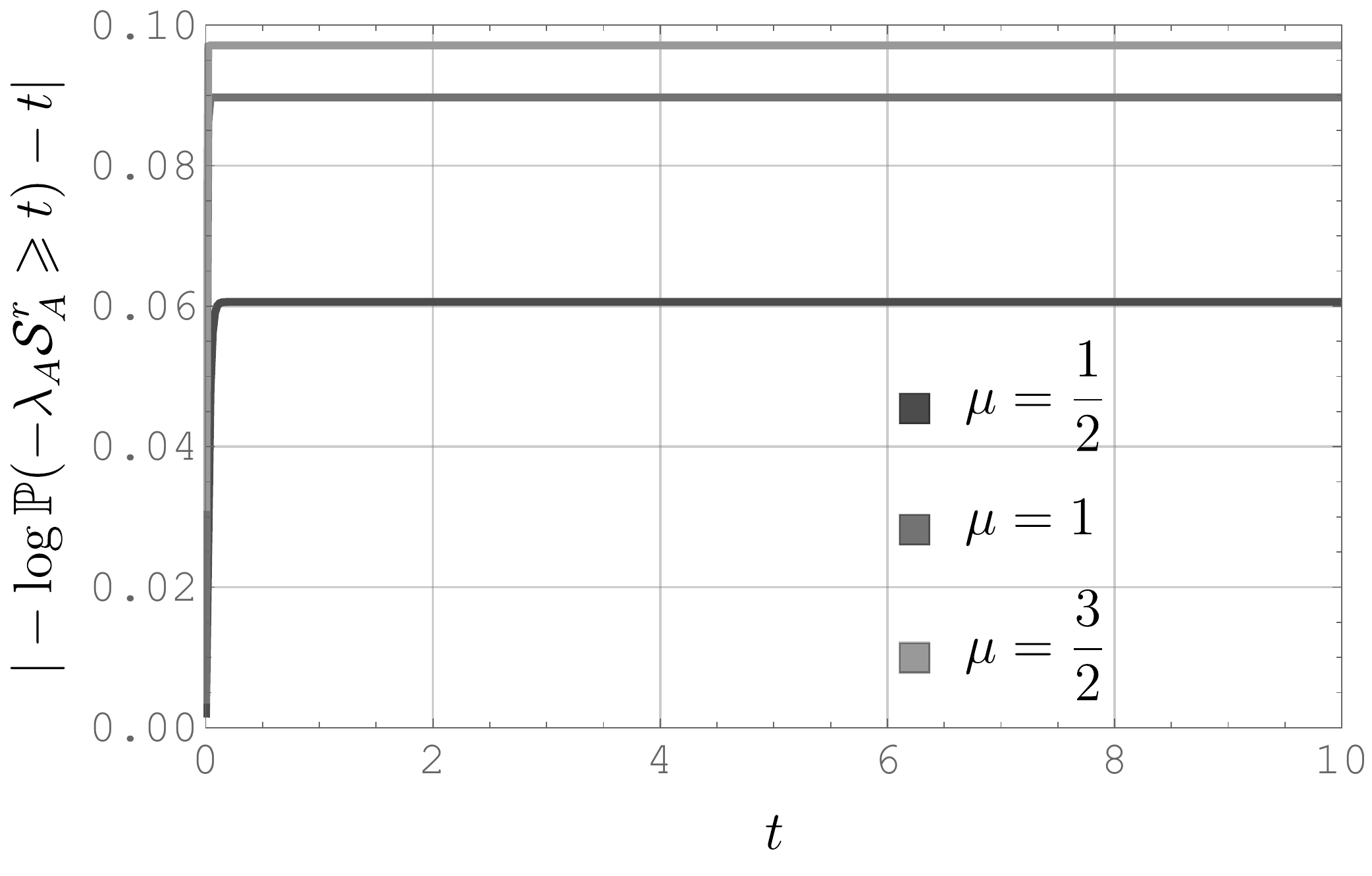}
        \caption{}
        \label{fig:logP-vs-t-A500-r50:abs-err}
    \end{subfigure}
    \caption{$\log\Pr(-\lambda_A\mathcal{S}_A^r\ge t)$ and $\abs{-\log\Pr(-\lambda_A\mathcal{S}_A^r\ge t)-t}$ as functions of $t$ for $t\in[0,10]$, $A=500$, $r=50$, and $\mu=\{1/2,1,3/2\}$.}
    \label{fig:logP-vs-t-A500-r50}
\end{figure}

\section{Concluding Remarks}
\label{sec:conclusion}

As was mentioned in the introduction, the obtained result, namely Theorem~\ref{thm:main-result} which we proved explicitly, is the continuous-time equivalent of a similar result obtained earlier by Pollak and Tartakovsky~\citeyearpar{Pollak+Tartakovsky:TPA2009} in the discrete-time setting; see also~\cite{Tartakovsky+etal:IWAP2008}. On a practical level, we were able to confirm experimentally that the GSR stopping time is approximately exponential even if the detection threshold is fairly low. Pollak and Tartakovsky~\citeyearpar{Pollak+Tartakovsky:TPA2009} made the same observation in the discrete-time setting. Although we already elaborated in the introduction on the significance of our findings to theoretical change-point detection, it is also worth adding that some of the new special functions identities utilized in the paper may prove useful in other areas as well, e.g., in stochastic processes, stochastic differential equations, mathematical physics, and mathematical finance, where special functions arise quite often.


\section*{Acknowledgement}
The author's effort was partially supported by the Simons Foundation via a Collaboration Grant in Mathematics under Award \#\,304574.


\end{document}